\documentclass[%
preprint,
 amsmath,amssymb,
 aps,
]{revtex4-2}
\usepackage{draftwatermark}
\SetWatermarkLightness{ 0.9 }
\SetWatermarkText{DRAFT}
\SetWatermarkScale{ 3 }
\usepackage{graphicx}
\usepackage{dcolumn}
\usepackage{bm}

\newcommand{\pfrac}[2]{\frac{\partial #1}{\partial #2}}
\newcommand{\pfracTwo}[2]{\frac{\partial^2 #1}{\partial #2^2}}
\newcommand{\yavg}[1]{\frac{1}{2b}\int\displaylimits^b_{-b} #1 dy}
\newcommand{\boldface}[1]{\boldsymbol{#1}}  

\newcommand{\bfn}{\boldface{n}}

\newcommand{\bft}{\boldface{t}}
\newcommand{\bfu}{\boldface{u}}

\newlength{\boxwidth}
\def\btheorem{\begin{theorem}}
\def\etheorem{\end{theorem}}
\def\blemma{\begin{lemma}}
\def\elemma{\end{lemma}}
\def\bproposition{\begin{proposition}}
\def\eproposition{\end{proposition}}
\def\bcorollary{\begin{corollary}}
\def\ecorollary{\end{corollary}}
\def\bdefinition{\begin{definition}}
\def\edefinition{\end{definition}}
\def\bexample{\begin{example}}
\def\eexample{\end{example}}
\def\bremark{\begin{remark}}
\def\eremark{\end{remark}}

\newcommand{\be}{\begin{equation*}}
\newcommand{\ee}{\end{equation*}}

\newcommand{\beq}{\begin{eqnarray*}}
\newcommand{\eeq}{\end{eqnarray*}}
\newcommand{\bem}{\begin{multline}}
\newcommand{\eem}{\end{multline}}
\newcommand{\ba}{\begin{align*}}
\newcommand{\ea}{\end{align*}}

\begin{document}


\title{Dissipation of nonlinear acoustic waves in thermoviscous pores}

\author{Krishna Sahithi}
\author{Prateek Gupta}%
 \email{prgupta@iitd.ac.in}
\affiliation{Department of Applied Mechanics\\
Indian Institute of Technology, Delhi,\\
Hauzkhas, New Delhi 110016\\
India
}%

\date{\today}

\begin{abstract}
We derive a nonlinear acoustic wave propagation model for analysing the thermoviscous dissipation in narrow pores with wavy walls. As the nonlinear waves propagate in the thermoviscous pores, the wave-steepening effect competes with the bulk dissipation, as well as the thermoviscous heat transfer and shear from the pore walls. Consequently, the length scale of the wave is modified. We use the characteristic nonlinear wave thickness scale to obtain linear and nonlinear wave equations governing the unsteady shock-wall interaction. We also perform two-dimensional shock-resolved DNS of the wave propagation inside the pores and compare the results with model equations. We show that for flat-walls and shock strength parameter $\epsilon$, the dimensional wall heat-flux and shear scale as $\epsilon$. For wavy walls, the scaling becomes $\epsilon^{3/2 - n(k)}$ where $k$ is the wall-waviness wavenumber and the exponent $n$ increases from $0.5$ for $k=0$ to $n(k)\approx0.65$ for $k=10$, $n(k)\approx 0.75$ for $k=20$, and $n(k)\approx0.85$ for $k=40$. Hence, increasing the wall waviness reduces the dependence of the wall heat-flux and shear on nonlinear acoustic wave strength. Furthermore, we show that both the dimensionless scaled wall shear and wall heat-flux decrease with increasing $k$.
\end{abstract}

\maketitle

\section{Introduction}
A nonlinear acoustic wave (or a shock wave) propagating in a narrow channel/pore results in a thermoviscous boundary layer close to the wall of the channel/pore~\cite{sichel1962leading}. This boundary layer results in the wall shear and wall heat-transfer in the fluid behind the wave which dissipate the wave. Such thermoviscous dissipation and the underlying boundary layer-unsteady shock interaction exist in a variety of technological applications such as thermoacoustics~\cite{rott1969damped, swift1988thermoacoustic, sugimoto_2016, gupta_lodato_scalo_2017},  aeroacoustics~\cite{tam2014experimental,zhang_bodony_2016,zhang2012numerical, kreitzman2024toward}, shock-tubes~\cite{sichel1962leading, ishida2014influence}, and detonation wave propagation~\cite{frederick2023statistical}. As the width/diameter of these channels/pores decreases, imperfections in the wall profile may have additional effects on the propagation and the dissipation of these strongly nonlinear acoustic waves. In this work, we study the effect of wavy walls on the dissipation and flow characteristics behind strongly nonlinear acoustic waves (weak shock waves) using two-dimensional direct numerical simulations (2D DNS) and a reduced order asymptotic model.
    
In general, high-amplitude or high-frequency acoustic waves exhibit nonlinear propagation~\cite{whitham2011linear}. In this work, we focus on the high-amplitude (or finite-amplitude) nonlinear acoustic waves, which steepen while propagating in a gas. Nonlinear steepening of acoustic waves is isentropic and leads to the formation of shock waves~\cite{gupta_lodato_scalo_2017, thirani2020knudsen,whitham2011linear}. As the shock-formation progresses, momentum and thermodynamic gradients steepen resulting in a thin region where non-isentropic processes due to thermoviscous dissipation are important. Tyagi and Sujith~\cite{tyagi2003nonlinear} showed that the steepening of nonlinear acoustic waves is adversely affected by varying cross-section and entropy gradients. While away from the shock waves, bulk thermoviscous effects are negligible, within the shock waves, thermoviscous dissipation balances the nonlinear steepening~\cite{gupta2018spectral} and is not negligible. This indicates towards an asymptotic-matching analysis, analogous to the boundary layer theory~\cite{white2006viscous}, possible for shock-dominated thermoviscous flows since the typical length scale (the shock thickness) depends on the local gradients the bulk thermoviscous processes. While the propagation of the nonlinear acoustic waves away from the walls has received a lot of attention, effects of wall shear and wall heat-transfer on nonlinear waves grazing past a rigid isothermal wall are less explored. As the waves propagate grazing a rigid wall, fluid particles close to the wall experience a phase lag in adjusting to the flow acceleration and temperature changes behind the wave~\cite{davies1969heat}. Such a lag can be quantified by the thermoviscous properties (viscosity and thermal conductivity) of the fluid. In a shock-tube, such unsteady shock-wall interaction results in the momentum and thermal boundary layers originating from the foot of the shock wave, causing attenuation of the strength of the shock wave~\cite{mirels1955laminar}. A few authors have investigated the effect of the induced boundary layer on the shock strength~\cite{duff1952interaction, emrich1953attenuation, bershader1956technical, hollyer1956attenuation, emrich1958wall}. Sichel~\cite{sichel1962leading} derived the analytical linearized expressions for shear stress and heat flux due to the induced boundary layers, assuming known strength of the shock wave grazing past a wall in a semi-infinite domain. In this work, we derive a combined asymptotic model for nonlinear acoustic waves propagating inside narrow thermoviscous two-dimensional pores.

 \begin{figure}[!t]
 \centering
 \includegraphics[width=1.0\textwidth]{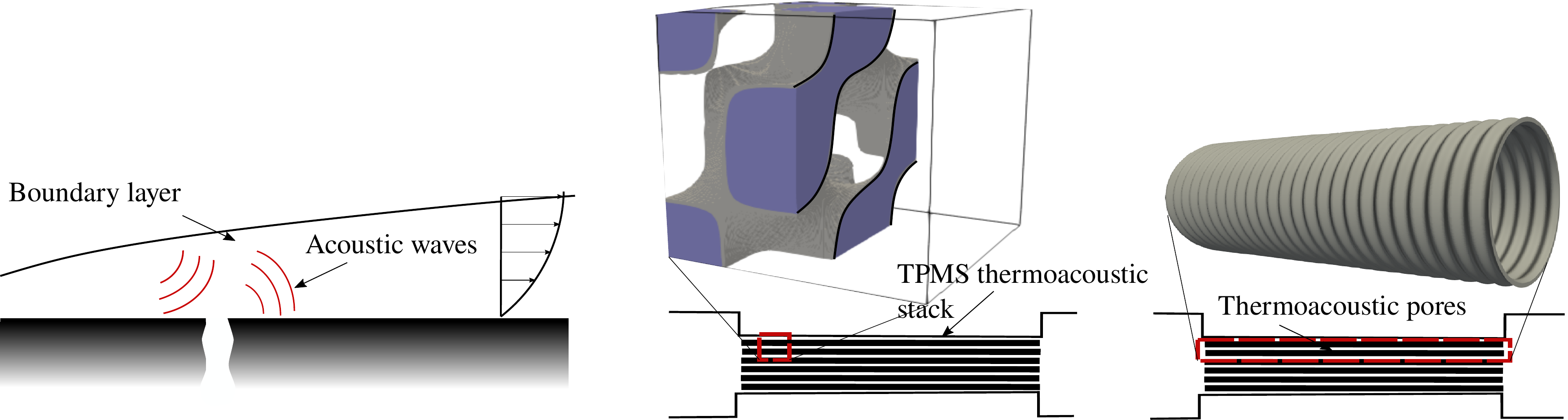}
 \put(-385,95){(a)}
  \put(-240,95){(b)}
 \put(-115,95){(c)}
 \caption{Schematic illustration of (a) an acoustically absorptive pore at the base of a high speed boundary layer formed on an otherwise flat plate, (b) stack of a thermoacoustic engine made with a TPMS heat exchanger (blue shaded region highlights the fluid), and (c) corrugations in a pore of a thermoacoustic stack.}
 \label{fig: intro}
\end{figure}
Shock waves grazing a rigid wall also exist in shock-tubes~\cite{spence1964review}. Usually, the diameter of shock-tubes is very large compared to the momentum and thermal boundary layers formed behind the wave. In narrow thermoviscous pores, the diameter and the boundary layer thickness (thermal and momentum) may be comparable~\cite{gupta_lodato_scalo_2017}.
Linear and nonlinear acoustic wave propagation in narrow thermoviscous pores plays an important role in a wide range of thermoacoustic and aeroacoustic applications~\cite{karpov2000nonlinear, patel2018impedance}. In thermoacoustics, spontaneous acoustic oscillations occur in resonators consisting of differentially heated \emph{stacks} which have narrow pores. Sugimoto and Shimizu~\cite{sugimoto2008boundary, shimizu2009physical, shimizu2010numerical, SugimotoShimizu_2024_JASA} and Sugimoto~\cite{sugimoto2010thermoacoustic, sugimoto_2016} have worked on the modelling of the linear thermoacoustic oscillations in very narrow and very wide limits of thermoviscous pores. In this work, we develop the nonlinear theory of wave-propagation in thermoviscous pores. In jet-engines, acoustic liners are used to regulate the noise generated during take-off and landing~\cite{hughes1990absorption, duff1952interaction,kirby1998impedance}. Furthermore, acousic liners and coatings with narrow pores are used along surfaces for delaying transition to turbulence in high-speed boundary layers~\cite{fedorov2003stabilization, patel2017towards}. In all these aeroacoustic applications, acoustic waves of a wide range of wavelengths interact with very narrow (compared to the typical flow length scales) pores inside the walls. A complete understanding of the thermoviscous dissipation of the acoustic waves inside these pores is pivotal for an effective design. Moreover, in all the aforementioned applications, having perfectly flat walls of the channels/pores is unfeasible owing to the manufacturing constraints. Additionally, a new class of materials known as TPMS (Triply Periodic Minimal Surfaces) are gaining attention due to analytically defined surfaces of a repeating unit cell for such applications~\cite{wang2016topological}.
Zhang et \emph{al.}~\cite{zhang2023analysis} and Lewis and Hickey~\cite{lewis2023conjugate} have analyzed the heat transfer in such materials with effusion cooling of the surfaces in a hypersonic flight as the focus. However, heat transfer and shear characteristics of such TPMS materials are largely unexplored so far. Due to the geometric features, TPMS unit cells have interconnected wavy pores through which a fluid or waves in the fluid can pass, thus making the study of effect of the waviness of walls on thermoviscous wave dissipation even more relevant. In figure~\ref{fig: intro} we summarize these relevant applications highlighting the importance of the thermoviscous dissipation of linear and nonlinear acoustic waves in narrow pores.

In this work, we develop a low order cross-sectionally averaged weakly nonlinear model that captures linear and nonlinear acoustic wave propagation in a thermoviscous pore (with flat walls or wavy walls) using the asymptotic analysis. We also draw comparisons between the model calculation and the DNS, conducted using a validated 2D fully compressible Navier-Stokes solver. We perform a parametric study by varying the strength of the nonlinear acoustic waves (shock waves) and the waviness of the walls of a 2D channel. In section~\ref{sec: theory}, we discuss the derivation of the model using the dimensionless governing equations. In section~\ref{sec: numerics}, we discuss the simulation setup for the 2D DNS and the parameteric space. In section~\ref{sec: results} we present the results on the variation of wall heat-flux and shear behind the propagating shock waves, comparing the results obtained from the weakly nonlinear model and the DNS throughout, before concluding in section~\ref{sec: conc}. Throughout, we refer to the 2D channels either as flat-walled pores or as wavy-walled pores, identifying that, in general, ``pore'' may refer to a cylindrical geometry. However, our modelling framework is independent of the Cartesian geometry used for calculations and simulations. Furthermore, we refer to the propagating nonlinear acoustic waves as shock waves.

\section{Theoretical model}
\label{sec: theory}
In this section, we discuss the modelling of the thermoviscous wave propagation and the flow induced behind the wave. We non-dimensionalize the governing equations using length scales appropriate to the propagation and diffusion of a shock wave. To this end, we consider the two-dimensional fully compressible Navier-Stokes equations governing the dynamics of a gas with no bulk viscosity,
\begin{align}
&\pfrac{\rho^*}{t^*} + \frac{\partial \left(\rho^* u^*\right)}{\partial x^*} + \frac{\partial \left(\rho^* v^*\right)}{\partial y^*} = 0,\label{eq: gov_rho}\\
&\rho^*\left(\pfrac{u^*}{t^*} + u^*\pfrac{u^*}{x^*} + v^*\pfrac{u^*}{y^*}\right) = -\pfrac{p^*}{x^*} + \frac{4}{3}\frac{\partial}{\partial x^*}\left(\mu\frac{\partial u^*}{\partial x^*}\right) + \frac{\partial}{\partial y^*}\left(\mu^*\frac{\partial u^*}{\partial y^*}\right) + \frac{\partial }{\partial y^*}\left(\mu^*\pfrac{v^*}{x^*}\right) \nonumber \\
&- \frac{2}{3}\frac{\partial }{\partial x^*}\left(\mu^*\pfrac{v^*}{y^*}\right),\label{eq: gov_rhou^*}
\end{align}
\begin{align}
&\rho^*\left(\pfrac{v^*}{t^*} + u^*\pfrac{v^*}{x^*} + v^*\pfrac{v^*}{y^*}\right) = -\pfrac{p^*}{y^*} + \frac{4}{3}\frac{\partial}{\partial y^*}\left(\mu^*\frac{\partial v^*}{\partial y^*}\right) + 
 \frac{\partial}{\partial x^*}\left(\mu^*\frac{\partial v^*}{\partial x^*}\right)+ \frac{\partial }{\partial x^*}\left(\mu^*\pfrac{u^*}{y^*}\right) \nonumber \\
&- \frac{2}{3}\frac{\partial }{\partial y^*}\left(\mu^*\pfrac{u^*}{x^*}\right),\label{eq: gov_rhov^*}\\
&\pfrac{(\rho^* e^*)}{t^*} + \pfrac{(\rho^* u^* e^*)}{x^*} + \pfrac{(\rho^* v^* e^*)}{y^*} = -p^*\left(\pfrac{u^*}{x^*} + \pfrac{v^*}{y^*}\right) +\nonumber\\& 2\mu^*\left(\left(\pfrac{u^*}{x^*}\right)^2 + \frac{1}{2}\left(\pfrac{u^*}{y^*} + \pfrac{v^*}{x^*}\right)^2 + \left(\pfrac{v^*}{y^*}\right)^2\right) - \frac{2\mu^*}{3}\left(\pfrac{u^*}{x^*} + \pfrac{v^*}{y^*}\right)^2 + \frac{\partial }{\partial x^*}\left(k^*\pfrac{T^*}{x^*}\right) +\nonumber\\& \frac{\partial}{\partial y^*}\left(k^*\pfrac{T^*}{y^*}\right),\label{eq: gov_rhoe} 
\end{align}
where $u^*$ and $v^*$ are the $x^*$ and $y^*$ direction velocities respectively, and $\rho^*, p^*, T^*, e^*, \mu^*,$ and $k^*$ denote the density, pressure, temperature, internal energy, dynamic viscosity, and thermal conduction coefficient of the gas, respectively. 
\begin{figure}[!b]
 \centering
 \includegraphics[width=0.8\textwidth]{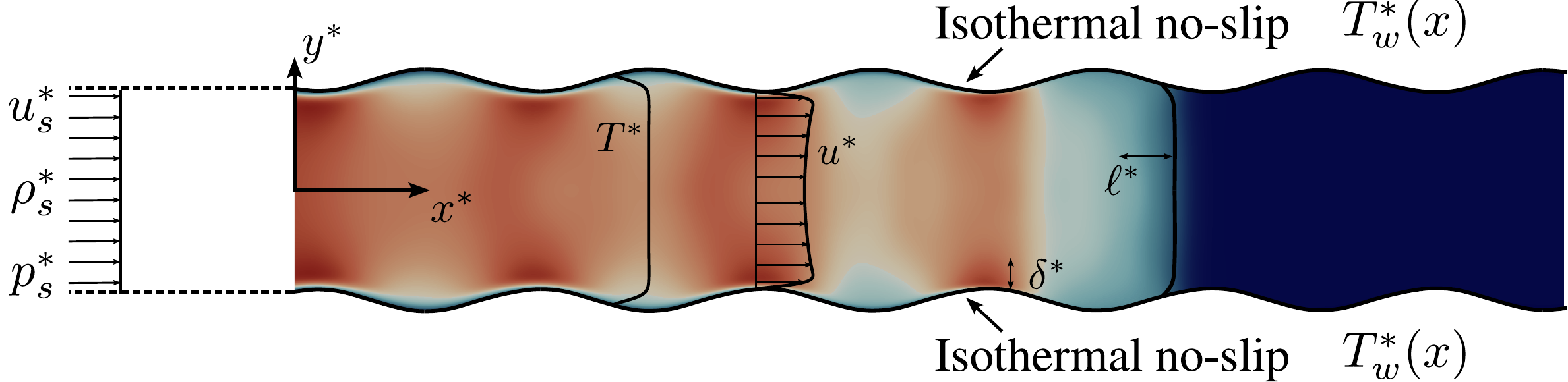}
 \caption{Schematic illustrating a nonlinear acoustic wave (a weak shock wave) propagating inside a narrow pore with wavy walls. The walls are modeled as isothermal no-slip walls with a prescribed temperature distribution of $T^*_w(x)$. Length scales in $x^*$ and $y^*$ direction are denoted by $\ell^*$ and $\delta^*$ respectively.}
\end{figure}
Throughout this work, the superscript $()^*$ denotes a dimensional quantity. We assume the gas to be ideal throughout our study, such that \eqref{eq: gov_rho}-\eqref{eq: gov_rhoe} are related via the perfect gas equations, 
\begin{equation}
 p^*=\rho^* R^*T^*,~~ e^* = c^*_v T^*= \frac{R^*T^*}{\gamma - 1}, \label{eq: perfect_gas}
\end{equation}
where $R^*$ is the gas constant, and $\gamma$ is the ratio of the specific heat at constant pressure $c^*_p$ to the specific heat at constant volume $c^*_v$. For viscosity and conduction coefficient, we assume the linear dependence on temperature (Rubesin law~\cite{white2006viscous}) with a constant Prandtl number, 
\begin{equation}
 \frac{\mu^*}{\mu^*_{\mathrm{ref}}} = C\frac{T^*}{T^*_{\mathrm{ref}}},~ \frac{\mu^* c^*_p}{k^*} = \frac{\mu^*_{\mathrm{ref}} c^*_p}{k^*_{\mathrm{ref}}} = \mathrm{Pr}.
\end{equation}
To non-dimensionalise the above equations, we consider the dimensionless variables, 
\begin{equation}
 p  = \frac{p^*}{\gamma p^*_0},~T  = \frac{T^*}{T^*_0},~\rho  = \frac{\rho^*}{\rho^*_0},~u  = \frac{u^*}{c^*_0},~v =\frac{v^*}{c^*_0},
\end{equation}
where $c^*_0 = \sqrt{\gamma R^* T^*_0}$ is the speed of sound at temperature $T^*_0$. For viscosity and conduction, we use the values at the mean temperature $T^*_0$ for scaling, 
\begin{equation}
 \mu  = \frac{\mu^*}{\mu^*_0} = T ,~k  = \frac{k^*}{k^*_0} = T .
\end{equation}
For $x^*$, we use a measure of the thickness of the shock wave ($\ell^*$) and for $y^*$, we use the thickness of the boundary layer behind the weak shock wave ($\delta^*$). For time, we use some time scale $\tau^*$, about which we will discuss further below. This yields, 
\begin{equation}
 x  = \frac{x^*}{\ell^*},~y =\frac{y^*}{\delta^*},~t  = \frac{t^*}{\tau^*}.
\end{equation}
Using the above scales, the dimensionless 2D equations become,
\begin{align}
 &\frac{\ell^*}{c^*_0\tau^*}\frac{\partial \rho }{\partial t } + \frac{\partial (\rho u )}{\partial x } + \frac{\ell^*}{\delta^*}\left(\frac{\partial (\rho v )}{\partial y }\right) = 0,\label{eq: nd_rho}\\
 &\rho \left(\frac{\ell^*}{\tau^* c^*_0}\frac{\partial u }{\partial t } + u \frac{\partial u }{\partial x } + \frac{\ell^*}{\delta^*}v \frac{\partial u }{\partial y }\right) =-\frac{\partial p }{\partial x } + \frac{1}{Re_{\ell^*}}\left(\frac{4}{3}\frac{\partial}{\partial x }\left(T \frac{\partial u }{\partial x }\right) + \left(\frac{\ell^*}{\delta^*}\right)^2\frac{\partial}{\partial y }\left(T \frac{\partial u }{\partial y }\right)\right)\nonumber\\
    &+ \frac{1}{Re_{\ell^*}}\left(\frac{\ell^*}{\delta^*}\right)\left(\frac{\partial}{\partial y }\left(T \frac{\partial v }{\partial x }\right)-\frac{2}{3}\frac{\partial}{\partial x }\left(T \frac{\partial v }{\partial y }\right)\right),\label{eq: nd_u} \\
 &\rho \left(\frac{\ell^*}{\tau^* c^*_0}\frac{\partial v }{\partial t } + \frac{\ell^*}{\delta^*}v \frac{\partial v }{\partial y } + u \frac{\partial v }{\partial x }\right) =-\frac{\ell^*}{\delta^*}\frac{\partial p }{\partial y } + \frac{1}{Re_{\ell^*}}\left(\frac{4}{3}\left(\frac{\ell^*}{\delta^*}\right)^2\frac{\partial }{\partial y }\left(T \frac{\partial v }{\partial y }\right) + \frac{\partial}{\partial x }\left(T \frac{\partial v }{\partial x }\right)\right)  \nonumber\\
    & + \frac{1}{Re_{\ell^*}}\left(\frac{\ell^*}{\delta^*}\right)\left(\frac{\partial}{\partial x }\left(T \frac{\partial u }{\partial y }\right)-\frac{2}{3}\frac{\partial}{\partial y }\left(T \frac{\partial u }{\partial x }\right)\right), \label{eq: nd_v}\\
 &\rho \left(\frac{\ell^*}{c^*_0\tau^*}\frac{\partial T }{\partial t }  + u \pfrac{T }{x } + \frac{\ell^*}{\delta^*}v \pfrac{T }{y }\right) + \gamma(\gamma -1)p \left(\pfrac{u }{x }+ \frac{\ell^*}{\delta^*}\pfrac{v }{y }\right) = \nonumber \\
  &\frac{\gamma}{Re_{\ell^*} \mathrm{Pr}}\left(\frac{\partial}{\partial x }\left(T \pfrac{T }{x }\right) + \frac{\ell^{*2}}{ \delta^{*2}}\frac{\partial }{\partial y }\left(T \pfrac{T }{y }\right)\right) + \nonumber \\
  &\frac{\gamma(\gamma - 1)}{Re_{\ell^*}}\left(2T \left[\left(\pfrac{u }{x }\right)^2 + \frac{1}{2}\left(\frac{\ell^*}{\delta^*}\pfrac{u }{y } + \pfrac{v }{x }\right)^2 + \frac{\ell^{*2}}{ \delta^{*2}}\left(\pfrac{v }{y }\right)^2\right] - \frac{2T }{3}\left(\pfrac{u }{x } + \frac{\ell^*}{\delta^*}\pfrac{v }{y }\right)^2\right). \label{eq: nd_T}   
\end{align}
The perfect gas equation reduces to the following dimensionless form, 
\begin{equation}
 \gamma p  = \rho T .\label{eq: nd_eos}
\end{equation}
Along with \eqref{eq: nd_rho}-\eqref{eq: nd_eos}, we also use the dimensionless governing equation for pressure, obtained by using \eqref{eq: gov_rhoe} and \eqref{eq: perfect_gas} given by, 
\begin{align}
 &\frac{\ell^*}{c^*_0\tau^*}\frac{\partial p }{\partial t }  + \pfrac{\left(p u \right)}{x } + \frac{\ell^*}{\delta^*}\pfrac{\left(p v \right)}{y } + (\gamma -1)p \left(\pfrac{u }{x }+ \frac{\ell^*}{\delta^*}\pfrac{v }{y }\right) = \nonumber \\
  &\frac{1}{Re_{\ell^*} \mathrm{Pr}}\left(\frac{\partial}{\partial x }\left(T \pfrac{T }{x }\right) + \frac{\ell^{*2}}{ \delta^{*2}}\frac{\partial }{\partial y }\left(T \pfrac{T }{y }\right)\right) + \nonumber \\
  &\frac{(\gamma - 1)}{Re_{\ell^*}}\left(2T \left[\left(\pfrac{u }{x }\right)^2 + \frac{1}{2}\left(\frac{\ell^*}{\delta^*}\pfrac{u }{y } + \pfrac{v }{x }\right)^2 + \frac{\ell^{*2}}{ \delta^{*2}}\left(\pfrac{v }{y }\right)^2\right] - \frac{2T }{3}\left(\pfrac{u }{x } + \frac{\ell^*}{\delta^*}\pfrac{v }{y }\right)^2\right).
  \label{eq: nd_p}
\end{align}
In~\eqref{eq: nd_u}-\eqref{eq: nd_p}, $Re_{\ell^*} = c^*_0\ell^*/\nu^*_0$ is the acoustic Reynolds number using the length scale $\ell^*$. We note that the bulk dissipation terms in \eqref{eq: nd_u} and \eqref{eq: nd_p} are linear in the flow variables if the variation of viscosity and conduction with temperature is ignored. Usually, in the derivation of nonlinear acoustic wave propagation equations~\cite{gupta_lodato_scalo_2017, gupta2018spectral, hamilton1998nonlinear}, the coordinate in the wave propagation direction ($x$ coordinate in this work) is scaled with the resonator length scale or the wavelength scale for a harmonic wave. Consequently, the bulk dissipation terms appear in both the linearized system of equations as well as the nonlinear equations in an \emph{ad hoc} manner. Ideally, the bulk dissipation terms are an order of magnitude smaller than the flow perturbation variables, as shown by Gupta and Scalo~\cite{gupta2018spectral}. To rectify this, we use the typical nonlinear acoustic wave length scale which results in a natural scaling of the bulk dissipation terms as terms of an order of magnitude smaller than the leading order. For nonlinear acoustic wave pulses, the typical length scale is governed by the bulk dissipation and the propagation speed of the wave (typical shock-thickness scale). Below, we discuss the derivation of the scaled nonlinear acoustics equations accounting for the wave thickness length scale as a typical scale in the $x$ direction.

\subsection{Scaling}
A typical thickness scale of a nonlinear acoustic wave (a weak shock wave) is $\nu^*_0/c^*_0$, which is a result of the nonlinear steepening balancing the bulk viscous dissipation~\cite{gupta2018spectral, jossy2023baroclinic}. To this end, we consider the length scale $\ell^*$ as, 
\begin{equation}
 \ell^* = \frac{\nu^*_0}{c^*_0\epsilon},\label{eq: x-scale}
\end{equation}
where $\epsilon$ is a small parameter denoting the strength of a weak shock wave. As $\epsilon$ increases from small values to $1$, the length scale approaches the shock-thickness scale. We note that Sichel~\cite{sichel1962leading} used the same scale to consider the linearized governing equations at the foot of a shock-wave grazing past a flat plate. Furthermore, the acoustic Reynolds number based on $\ell^*$ becomes, 
\begin{equation}
 Re_{\ell^*} = \frac{1}{\epsilon}.\label{eq: acoustic_Re_l}
\end{equation}
Additionally, we use the time scale $\tau^*$ as a typical time scale of the acoustic fluctuations over the length $\ell^*$, 
\begin{equation}
 \tau^* = \frac{\ell^*}{c^*_0}.\label{eq: t-scale}
\end{equation}
With the above choice of the length scale $\ell^*$ and the time scale $\tau^*$, we assume the following perturbation expansions in $p,\rho,T,u,$ and $v$,
\begin{equation}
 p = \frac{1}{\gamma} + \epsilon p',~\rho = f(x) + \epsilon \rho',~T = h(x) + \epsilon T',~ u = \epsilon u',~ v=\epsilon v'.
\label{eq: perturbation_expansions}
\end{equation}
Note that we assume an isobaric temperature distribution along $x$ in the base state since most of the applications of the current work exhibit length-wise temperature gradients (in thermoacoustics, the direct external heating and in aeroacoustics, the aerothermodynamic heating). The continuity equation using the perturbation expansion yields, 
\begin{equation}
 \frac{\partial \rho'}{\partial t} + \pfrac{(f u')}{x} + \epsilon\pfrac{(\rho'u')}{x} + \frac{f}{Re_{\delta^*} \epsilon}\pfrac{v'}{y} + \frac{1}{Re_{\delta^*}}\pfrac{(\rho'v')}{y} = 0.\label{eq: pert_cont_1}
\end{equation}
Furthermore, the $x$ velocity equation yields, 
\begin{align}
&f\left(\frac{\partial u'}{\partial t}\right) + \epsilon\left(\rho'\pfrac{u'}{t} + f u'\pfrac{u'}{x}\right) + \frac{f }{Re_{\delta^*}}v'\pfrac{u'}{y} = -\pfrac{p'}{x}  + 
 \epsilon\left(\frac{4}{3}\frac{\partial}{\partial x}\left(h\pfrac{u'}{x}\right) + \epsilon\frac{\partial}{\partial x}\left(T'\pfrac{u'}{x}\right)\right)+\nonumber \\
 &\frac{1}{Re^2_{\delta^*} \epsilon} \left(h\pfracTwo{u'}{y} + \epsilon\frac{\partial}{\partial y}\left(T'\pfrac{u'}{y}\right) \right)+
 \frac{1}{Re_{\delta^*}}\left(\frac{\partial}{\partial y}\left(h\pfrac{v'}{x}\right) - \frac{2}{3}\frac{\partial }{\partial x}\left(h\pfrac{v'}{y}\right)\right)+\nonumber\\
 &\frac{\epsilon}{Re_{\delta^*}}\left(\frac{\partial}{\partial y}\left(T'\pfrac{v'}{x}\right) - \frac{2}{3}\frac{\partial }{\partial x}\left(T'\pfrac{v'}{y}\right)\right).\label{eq: pert_u_1}
\end{align}
To capture the flow physics close to the shock-wall interaction, capturing the $y$ direction velocity is essential. Far away from both the walls and the shock wave, $v'\ll u'$ holds. Moreover, gradients of $v'$ in all directions are also very small. For wavy walls, the gradients in $v'$ are proportional to the local slope of the walls. However, near the interaction zone between the shock wave and the wall, terms consisting of the gradients of $v'$ in the continuity equation can not be negligible since $v'$ vanishes at the wall. For the gradients of $v'$ to influence the continuity \eqref{eq: pert_cont_1} and for shear stress effects to influence the $x$ velocity at the leading order, the following scaling relations must hold,
\begin{equation}
 \frac{|v'|}{Re_{\delta^*}} \sim \epsilon,~~Re_{\delta^*} = \epsilon^{-1/2}.\label{eq: scalings}
\end{equation}
The above scaling relations show that the length scale ratio is given by, 
\begin{equation}
 \frac{\delta^*}{\ell^*} = \epsilon^{1/2}.\label{eq: length_scale_ratio}
\end{equation}
Hence for very small $\epsilon$, the typical $y$ direction length scale is very small compared to the $x$ direction length scale. As $\epsilon$ increases, the wave thickness and the boundary layer thickness behind the wave become more comparable. Below, we consider equations correct up to $\mathcal{O}(1)$ (linear) and $\mathcal{O}(\epsilon)$ (nonlinear). The linearized equations are identical to the standard thermoacoustic equations for acoustic waves. However, the nonlinear equations are novel in that the dissipation and nonlinear terms are scaled with $\epsilon$ with no \emph{ad hoc} assumptions regarding the length scales. We integrate the nonlinear equations in $y$ direction to obtain the cross-sectionally averaged model equations.


\subsection{Linear model}

Using the perturbation expansions for $v'$ as, 
\begin{equation}
 v' = \epsilon^{1/2} v^{(1)},
\end{equation}
and for $u', p', \rho', T'$ as, 
\begin{equation}
    (u', p', T', \rho') = \left(u^{(1)}, p^{(1)}, T^{(1)}, \rho^{(1)}\right) + \epsilon \left(u^{(2)}, p^{(2)}, T^{(2)}, \rho^{(2)}\right)
\end{equation}
and the scale relations in \eqref{eq: scalings}, we obtain the following governing equations at the leading order 
\begin{align}
 &\frac{\partial \rho^{(1)}}{\partial t} + \frac{\partial (fu^{(1)})}{\partial x} + f\frac{\partial v^{(1)}}{\partial y} = 0,\label{eq: cont_nd_o1}\\
 &f\frac{\partial u^{(1)}}{\partial t} + \pfrac{p^{(1)}}{x} = h\pfracTwo{u^{(1)}}{y},\label{eq: x_nd_o1}\\
 &\pfrac{p^{(1)}}{y} = 0,\label{eq: y_nd_o1}\\
 &f\left(\pfrac{T^{(1)}}{t} + u^{(1)}\frac{dh}{dx}\right) + (\gamma - 1)\left(\pfrac{u^{(1)}}{x} + \pfrac{v^{(1)}}{y}\right) = \frac{\gamma h}{\mathrm{Pr}}\pfracTwo{T^{(1)}}{y},\label{eq: T1_nd_o1}
\end{align}
combined with the linear perfect gas equation, 
\begin{equation}
    \rho^{(1)} h + T^{(1)}f = \gamma p^{(1)}.~\label{eq: eos_nd_o1}
\end{equation}
Using \eqref{eq: cont_nd_o1},~\eqref{eq: T1_nd_o1},~and~\eqref{eq: eos_nd_o1}, we obtain the dimensionless equation for temperature perturbation at the leading order as,
\begin{equation}
 f\left(\pfrac{T^{(1)}}{t} + u^{(1)}\frac{dh}{dx}\right) - (\gamma - 1)\pfrac{p^{(1)}}{t} = \frac{h}{\mathrm{Pr}}\pfracTwo{T^{(1)}}{y}.~\label{eq: T_nd_o1}
\end{equation}
Using the isothermal no-slip boundary conditions 
\begin{equation}
 u^{(1)}(x,\pm b,t) = 0,~\mathrm{and}~~T^{(1)}(x,\pm b, t) = 0,
\end{equation}
\eqref{eq: x_nd_o1} and \eqref{eq: T_nd_o1} can be solved using the Laplace transform, yielding $u^{(1)}$ and $T^{(1)}$ as functions of $p^{(1)}$ as, 
\begin{align}
&u^{(1)} = \frac{1}{f}\int^t_0\mathcal{G}(t-\tau, 1/h)\pfrac{p^{(1)}}{x}(\tau)d\tau + \mathcal{O}(\epsilon),\label{eq: u_sol_lapl}\\
&\frac{\partial T^{(1)}}{\partial t} = K\int^t_0\mathcal{G}(t-\tau, \sqrt{\mathrm{Pr}}/h)\pfrac{p^{(1)}}{x}(\tau)d\tau + (\gamma - 1)h\int^t_0\mathcal{G}(t-\tau, \sqrt{\mathrm{Pr}}/h)\pfracTwo{p^{(1)}}{t}(\tau)d\tau \nonumber\\
 &-\mathrm{Pr}K \int^t_0\mathcal{G}(t-\tau, 1/h)\pfrac{p^{(1)}}{x}(\tau)d\tau+ \mathcal{O}(\epsilon),\label{eq: T_sol_lapl}
\end{align}
where the kernel $\mathcal{G}(t,a)$ is given by, 
\begin{equation}
 \mathcal{G}(t,a) = 2\sum^{\infty}_{n=0}\left(-1\right)^{n+1}\frac{e^{\frac{-z^2_n t}{ab}}}{z_n}\cos\left(z_n \frac{y}{b}\right)~~\mathrm{where}~~ z_n = \left(n + \frac{1}{2}\right)\pi,\label{eq: relaxation_kernel}
\end{equation}
and $K=\frac{h}{1-\mathrm{Pr}}\frac{dh}{dx}$. Integration of~\eqref{eq: cont_nd_o1}-~\eqref{eq: T_nd_o1} yields the cross-sectionally averaged equations as,
\begin{align}
 &f\pfrac{\overline{u}^{(1)}}{t} + \left(\frac{\partial }{\partial x} + \frac{1}{b}\frac{db}{dx}\right)\overline{p}^{(1)} = \frac{h\tau_w}{b},\\
 &\pfrac{\overline{p}^{(1)}}{t} +  \left(\frac{\partial }{\partial x} + \frac{1}{b}\frac{db}{dx}\right)\overline{u}^{(1)} = \frac{hq_w}{b\mathrm{Pr}},
\end{align}
where $\overline{\left(\right)}$ denotes the average in $y$ direction and $\tau_w$ and $q_w$ are the wall shear-stress and wall heat-transfer. Equation \eqref{eq: relaxation_kernel} also shows that the eigenfunction expansion method can be used to obtain identical results~\cite{gupta_lodato_scalo_2017}. Moreover, pressure perturbations at the leading order are uniform in $y$. Hence, $v^{(1)}$ can be calculated using the continuity \eqref{eq: cont_nd_o1} as, 
\begin{equation}
 v^{(1)} = \left(\left(\frac{\partial }{\partial x} + \frac{1}{b}\frac{db}{dx}\right)\overline{u}^{(1)} - \frac{hq_w}{bPr}\right)y + \frac{h}{\mathrm{Pr}}\pfrac{T^{(1)}}{y} - \frac{\partial }{\partial x}\int\displaylimits^y_0 u^{(1)}dy.
 \label{eq: v_1}
\end{equation}
We note that in the linear model, the bulk thermoviscous dissipation terms $\partial^2 u'/\partial x^2$ and $\partial^2 T'/\partial x^2$ terms do not appear due to the correct scaling of the $x$ coordinate. This also proves that even though these terms are seemingly linear in perturbation variables $u'$ and $T'$, upon correct scaling, they are of higher order (shown computationally in \cite{gupta2018spectral}).
Additionally, \eqref{eq: v_1} captures the $y$ direction velocity due to the thermoviscous wave propagation in the $x$ direction. For wavy walls, the propagation in $x$ direction excites $y$ direction acoustic modes as well, which are not captured in~\eqref{eq: v_1} since the $y$ direction momentum equation yields a trivial result~\eqref{eq: y_nd_o1}, similar to the boundary layer theory.
\subsection{Nonlinear model}
The linearized or $\mathcal{O}(1)$ equations~\eqref{eq: cont_nd_o1}-\eqref{eq: T1_nd_o1} are identical to the time-domain linear thermoacoustic equations~\cite{gupta_lodato_scalo_2017}. These hold only for very long wavelength perturbations, that is, for perturbations of wavelength much longer than the shock thickness scale. For relatively sharper nonlinear acoustic waves, the governing equations correct up to $\mathcal{O}(\epsilon)$ are given by,
\begin{align}
 f\pfrac{u'}{t} + \epsilon\left(fu'\pfrac{u'}{x} + v'\pfrac{u'}{y}\right) = -\left(1 - \frac{\epsilon \rho'}{f}\right)\pfrac{p'}{x} + \left(1 - \frac{\epsilon \rho'}{f}\right)h\pfracTwo{u'}{y} + \nonumber\\
 \epsilon\left[\frac{4}{3}\frac{\partial}{\partial x}\left(h\pfrac{u'}{x}\right) + \frac{\partial}{\partial y}\left(T'\pfrac{u'}{y}\right) + \frac{\partial }{\partial y}\left(\frac{h}{3}\pfrac{v'}{x}\right) - \frac{2}{3}\frac{dh}{dx}\pfrac{v'}{y}\right],
 \label{eq: up_nl}
\end{align}
and 
\begin{align}
  \pfrac{p'}{t} + \left(\pfrac{u'}{x} + \pfrac{v'}{y}\right) + \epsilon\left(u'\pfrac{p'}{x} + v'\pfrac{p'}{y} + \gamma p'\left(\pfrac{u'}{x} + \pfrac{v'}{y}\right)\right) = \frac{h}{\mathrm{Pr}}\pfracTwo{T'}{y} + \nonumber\\
 \frac{\epsilon}{\mathrm{Pr}}\left(\pfracTwo{(hT')}{x} + \frac{1}{2}\pfracTwo{(T'^2)}{y} + Pr(\gamma - 1)h\left(\pfrac{u'}{y}\right)^2\right).~\label{eq: pp_nl}
\end{align}
For wave pulses of smaller length scales or typical nonlinear acoustic waves, we derive the $y-$averaged equations correct up to $\mathcal{O}(\epsilon)$ as,
\begin{align}
 f\pfrac{\overline{u}'}{t} + \yavg{\epsilon f\left(u'\pfrac{u'}{x} + v'\pfrac{u'}{y}\right)} = -\left(\pfrac{}{x} + \frac{1}{b}\frac{db}{dx}\right)\overline{p}' + \frac{\epsilon\overline{\rho}}{f}\pfrac{\overline{p}'}{x} + \frac{h\tau_w}{b}\left(1 - \epsilon\gamma \overline{p}'\right) \nonumber \\
 - \epsilon\yavg{\pfrac{T'}{y}\pfrac{u'}{y}} + \frac{4\epsilon}{3}\left(\frac{\partial }{\partial x} + \frac{1}{b}\frac{db}{dx}\right)\left(h\left(\pfrac{}{x} + \frac{h}{b}\frac{db}{dx}\right)\overline{u}'\right),\label{eq: u_bar_nl}
\end{align}
and
\begin{align}
 \pfrac{\overline{p}'}{t} + \left(\pfrac{}{x} + \frac{1}{b}\frac{db}{dx}\right)\overline{u}' + \epsilon\left(\gamma \overline{p}'\left(\pfrac{}{x} + \frac{1}{b}\frac{db}{dx}\right)\overline{u}' + \overline{u}\pfrac{\overline{p}'}{x} \right) = \frac{hq_w}{b Pr} + \frac{\epsilon}{\mathrm{Pr}}\left(\frac{\partial }{\partial x} + \frac{1}{b}\frac{db}{dx}\right)^2 \left(h\overline{T}'\right) \nonumber\\
 +\epsilon h(\gamma - 1)\yavg{\left(\pfrac{u'}{y}\right)^2}.\label{eq: p_bar_nl}
\end{align}
In section~\ref{sec: results}, we discuss the numerical results obtained using the 2D DNS (see section~\ref{sec: numerics}) and time integration of \eqref{eq: u_bar_nl} and \eqref{eq: p_bar_nl}. We note that we need $q_w$ and $\tau_w$ in \eqref{eq: u_bar_nl} and \eqref{eq: p_bar_nl} to solve for $\overline{p}'$ and $\overline{u}'$. One choice is to solve for the full nonlinear $u'(x,y,t)$ and $T'(x,y,t)$ fields and compute the wall shear and wall heat-flux. However, that beats the purpose of developing a low order model. To this end, we use the linearized solution in \eqref{eq: u_sol_lapl} and \eqref{eq: T_sol_lapl} to compute the instantaneous $\tau_w(x,t)$ and $q_w(x,t)$ at every step. Additionally, we use the linearized solution in all the terms involving products of fields varying in $y$ direction. In~\eqref{eq: u_sol_lapl} and \eqref{eq: T_sol_lapl} we use the pressure $\overline{p}'$ obtained from the nonlinear equations to avoid solving for another set of leading order equations at every time step. 

It is remarkable that in the long-wavelength limit (wavelength longer than shock thickness scale), we obtain the usual linearized thermoacoustic wave equations~\cite{sugimoto2010thermoacoustic, gupta_lodato_scalo_2017, rott1969damped}, usually derived without any formal scaling analysis. As the wavelength scale decreases and the amplitude increases (increasing $\epsilon$),~\eqref{eq: u_bar_nl} and~\eqref{eq: p_bar_nl} become more relevant. A detailed derivation of \eqref{eq: u_bar_nl} and~\eqref{eq: p_bar_nl} utilizing~\eqref{eq: up_nl} and~\eqref{eq: pp_nl} is given in  appendix~\ref{sec: Appb}. Below, we discuss the computational setup used for the 2D DNS and numerical solution of the model equations.
\begin{figure}[!b]
 \centering
 \includegraphics[width=0.75\textwidth]{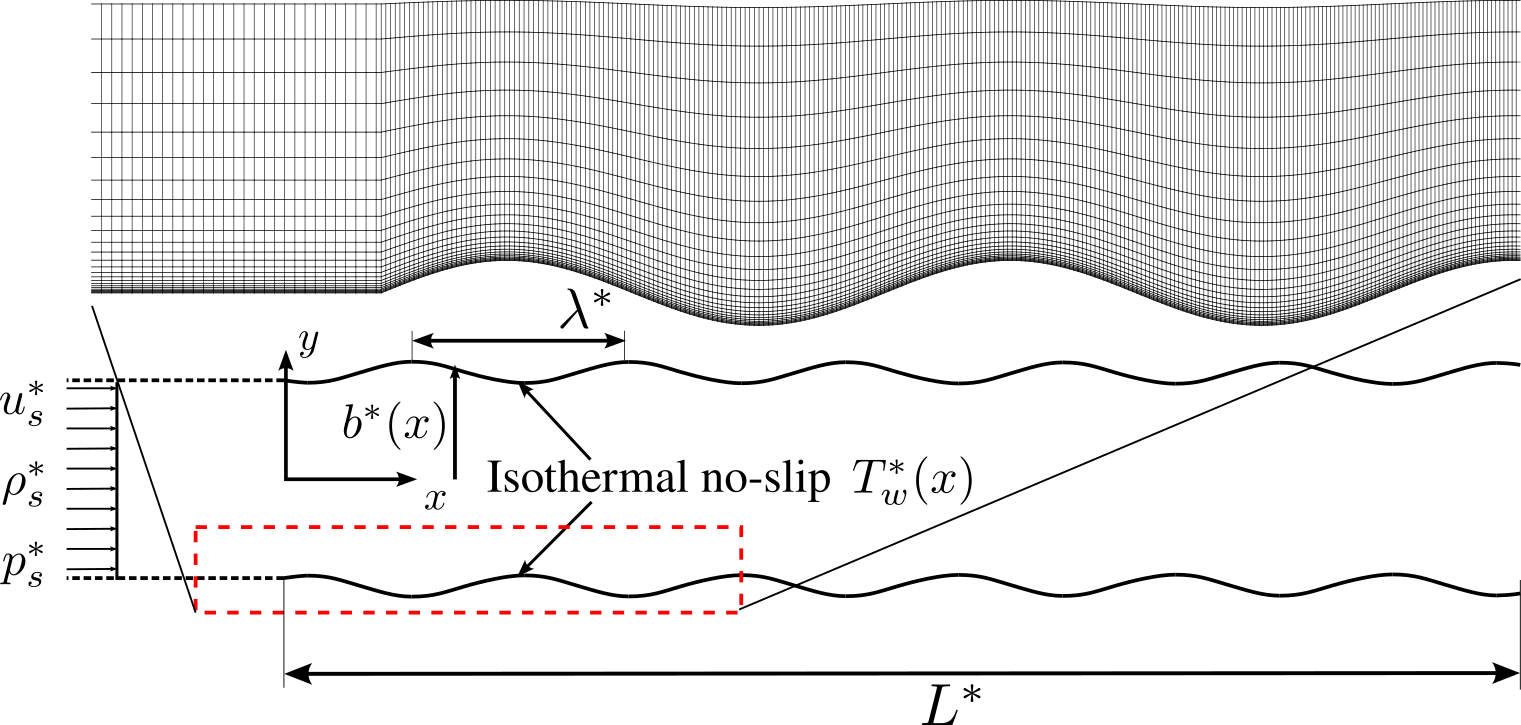}
 \caption{Schematic of a zoomed in section of the quadrilateral element mesh used for the 2D DNS. $\lambda^*$ denotes the dimensional wavelength of the walls and $b^*(x)$ is the half-width of the viscous pore. }
 \label{fig: wavy_mesh}
\end{figure}

\section{Numerical Setup}
\label{sec: numerics}
For the two-dimensional direct numerical simulations (2D DNS), we use an in-house C++ based spectral difference solver for fully compressible Navier-Stokes equations on unstructured quadrilateral grids. The solver is validated against the self-similar solution to the steady high-speed flat plate laminar boundary layers (see supplementary information). Since the primary goal of this study is to capture and model the effect of the waviness of the walls on the thermoviscous wave dissipation, we consider four values of the wall-waviness wavenumber $k=L^*/\lambda^*$ (see figure~\ref{fig: wavy_mesh}) starting from 0 (flat wall). For each $k$, we consider three values of the shock-strength parameter $\epsilon$. For each value of $\epsilon$, we compute the steep jump in pressure, velocity, and temperature using the normal shock Mach number $M_s = 1+\epsilon$ ($u^*_s, p^*_s, \rho^*_s$ in figure~\ref{fig: wavy_mesh}) in a duct section of length $1\times 10^{-3}~\mathrm{m}$ with straight and inviscid walls. The initial dimensional pressure perturbation is given by,
\begin{equation}
 p'^* = p^*_s(\epsilon)\left(\frac{1-\tanh\left(\alpha(x^*-x^*_0)\right)}{2}\right),
 \label{eq: sim_init_p}
\end{equation}
where $p^*_s(\epsilon)$ denotes the post-shock dimensional pressure for a normal shock with Mach number $M_s = 1+\epsilon$. To initialize only the forward propagating Riemann invariant, velocity and density in the inviscid-wall duct section are initialized as~\cite{gupta2018spectral}, 
\begin{equation}
\rho^*_s = \rho^*_0\left(\frac{p^*_0 + p'^*}{p^*_0}\right)^{1/\gamma},~u^* = \frac{2c^*_0}{\gamma - 1}\left(\left(\frac{p^*_0 + p'^*}{p^*_0}\right)^{(\gamma - 1)/2\gamma} - 1\right)
\label{eq: sim_init_urho}
\end{equation}
For all the simulations, we take $\alpha=2.5\times 10^4~\mathrm{m}^{-1}$ and $x^*_0=5\times 10^{-4}~\mathrm{m}$. Equal values of $\alpha$ are considered for all the cases highlighting that the thickness of the waves as they enter the viscous region are similar (not identical), since the inviscid duct section is not long enough to allow the waves to reach the natural weak-shock thickness (which would be very long given the shocks are weak and hence would increase the computational cost tremendously). We note that the values obtained from \eqref{eq: sim_init_p} and~\eqref{eq: sim_init_urho} are obtained assuming isentropic changes in the base state. As the initial perturbation propagates in the inviscid-wall, the shock wave develops before entering the viscous thermoacoustic pore region of length $0.02~\mathrm{m}$ and width $2b^*(x)$ with no-slip isothermal walls. The function $b^*(x)$ describes the wall geometry. To study the effect of wall waviness on propagation of nonlinear acoustic wave dissipation, we take $b^*(x)$ as, 
\begin{equation}
 b^*(x) = \left(\frac{1}{2}  - 0.05\sin\left(\frac{2\pi k x^*}{L^*}\right)\right)~\mathrm{mm},
 \label{eq: wavy}
\end{equation}
where $k = L^*/\lambda^*$ is the wavenumber of the wall geometry (c.f. figure~\ref{fig: wavy_mesh} for $\lambda^*$). We vary $k$ from $0$ to $40$ in increments of $10$ (see table~\ref{tab: simulation_cases}). We impose temperature $T^*_w(x)$ on the no-slip isothermal walls to simulate thermoacoustic effects. In all our simulations, we assume linear variation of the dimensionless wall temperature $T_w(x)$ as, 
\begin{equation}
 T_w(x) = \frac{1}{T^*_0}\left(T^*_0 + \left(\frac{T^*_H - T^*_0}{L^*}\right)x^*\right) = 1 + \frac{T^*_H/T^*_0 - 1}{L^*}x^*,
\end{equation}
with $T^*_H/T^*_0 = 2$. 

\begin{table}[!t]
\centering
\caption{Simulation cases and the corresponding shock-strength parameter $\epsilon$, scaled wall wavenumber $k = L^*/\lambda^*$ (c.f.~\eqref{eq: wavy}), maximum $\Delta y^+$, and maximum $\Delta x = \Delta x^*/\ell^*$.}
\label{tab: simulation_cases}
 \begin{tabular}{c|c|c|c|c}
 \hline
 Case &  $\epsilon$ & $k$ & $\Delta y^+_{\mathrm{max}}$ & $\Delta x_{\mathrm{max}}$\\
 \hline
 $1a$ & $1\times 10^{-3}$ & 0 & 0.034 & 0.122\\
 $1b$ & $1\times 10^{-3}$ & 10 & 0.033 & 0.122\\
 $1c$ & $1\times 10^{-3}$ & 20 & 0.030 & 0.122\\
 $1d$ & $1\times 10^{-3}$ & 40 & 0.055 & 0.122\\
 $2a$ & $5\times 10^{-4}$ & 0 & 0.023 & 0.061\\
 $2b$ & $5\times 10^{-4}$ & 10 & 0.023 & 0.061\\
 $2c$ & $5\times 10^{-4}$ & 20 &0.021 & 0.061\\
 $2d$ & $5\times 10^{-4}$ & 40 & 0.040 & 0.061\\
 $3a$ & $2\times 10^{-3}$ & 0 & 0.048 & 0.245\\
 $3b$ & $2\times 10^{-3}$ & 10 & 0.047& 0.245\\
 $3c$ & $2\times 10^{-3}$ & 20 & 0.042& 0.245\\
 $3d$ & $2\times 10^{-3}$ & 40 & 0.079& 0.245\\
  \hline
 \end{tabular}
\end{table}

As the wave enters the viscous region, the shear stress and heat transfer at the no-slip isothermal walls peak, resulting in the wave dissipation. To assess the resolved nature of the simulations we report maximum $\Delta y^+$ for all the simulation cases right next to the walls calculated as, 
\begin{equation}
 \Delta y^+_{\mathrm{max}} = \Delta y^*\sqrt{\frac{1}{\nu^*_0}\left(\frac{\partial u^*}{\partial y^*}\right)\Big|_{\mathrm{max}}}.
\end{equation}
in table~\ref{tab: simulation_cases}. Since $\Delta y^+ < 0.1$ for all the simulations, the momentum boundary layer is well resolved in all the cases. We use $\mathrm{Pr}=0.72$ in all of the simulations. Hence, the thermal boundary layer is thicker than the momentum boundary layer and is naturally resolved. Furthermore, the maximum value of the dimenionless $x$ resolution, $\Delta x$, is also reported in table~\ref{tab: simulation_cases}. For $\Delta x_{\mathrm{max}} < 1$, shock waves are resolved. We note that both $\Delta y^+$ and $\Delta x$ correspond to the size of the elements in the mesh. We use our spectral-difference solver with three Gauss-Legendre collocation points in each direction in each cell, which corresponds to a second order polynomial reconstruction in each cell. Hence, the effective resolution within each cell is smaller than $\Delta y^+$ and $\Delta x$ in $y$ and $x$ directions, respectively.

For the low order model equations~\eqref{eq: u_bar_nl} and \eqref{eq: p_bar_nl}, we use the second-order central-difference scheme with staggered grid for $\overline{p}'$ and $\overline{u}'$ in space and RK4 time integration \cite{gupta_lodato_scalo_2017}. The model equations are integrated in dimensionless space and time coordinates. In the section below, we compare the DNS and model equation results and elucidate the dissipation mechanism of waves in the wavy wall thermoviscous pores.

\section{Results and Discussion}
\label{sec: results}

\begin{figure}[!b]
 \centering
 \includegraphics[width=\textwidth]{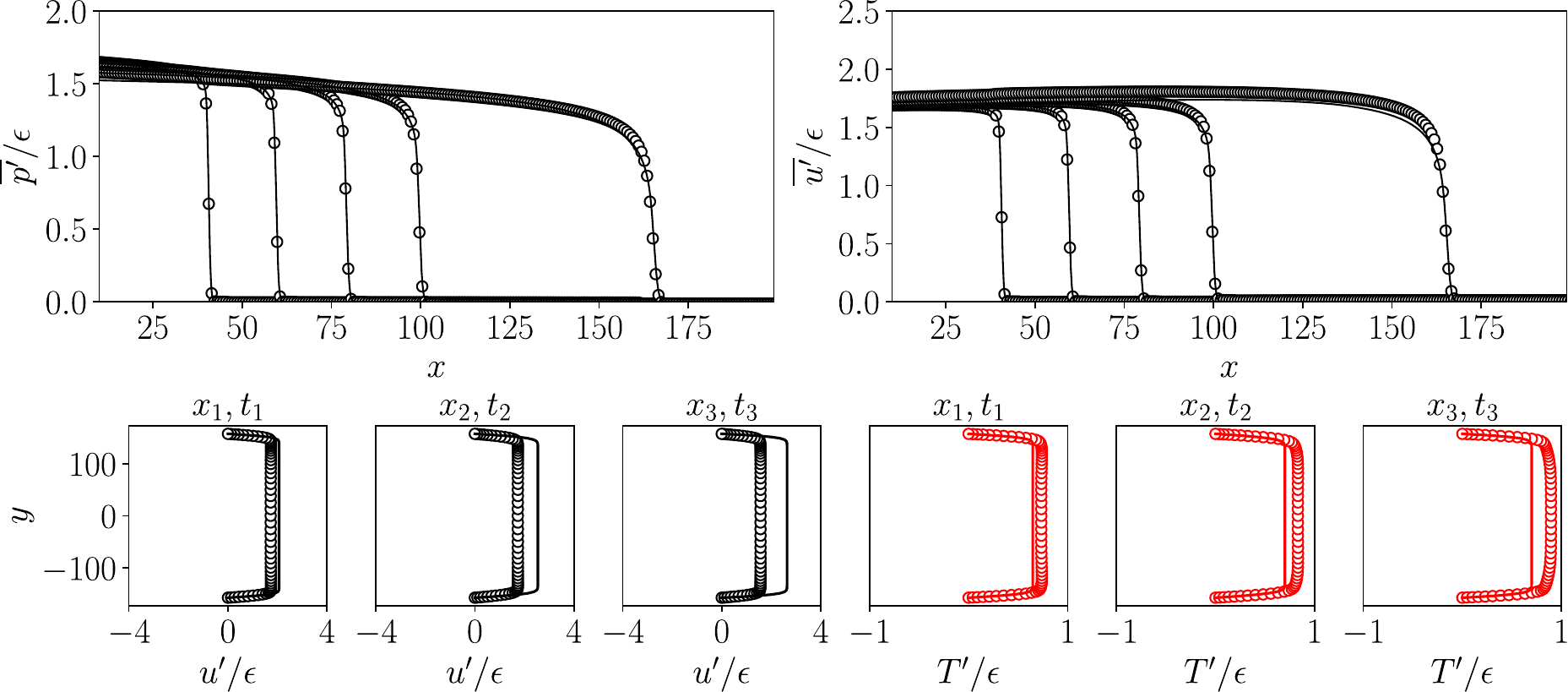}
  \put(-385,170){$(a)$}
 \put(-190,170){$(b)$}
  \put(-385,70){$(c)$}
 \caption{Comparison of the $y$-averaged dimensionless $(a)$ pressure and $(b)$ velocity perturbation profiles in $x$ obtained from the nonlinear model equations (lines) with the DNS data (markers) at dimensionless times $t = 34.5, 51.8, 69.1, 86.4, 138.2$. $(c)$ Comparison of the dimensionless velocity and temperature profiles in $y$ at $x_1=50, x_2=110, x_3=180$ at times $t_1=51.8, t_2=105.3, t_3=157.2$ obtained from the nonlinear model equations (lines) with DNS (markers). All $(a),(b)$, and $(c)$ correspond to the case $1a$ in Table~\ref{tab: simulation_cases}.}
 \label{fig: flat_pore_y-avgd}
\end{figure}

In this section, we present the results of the 2D DNS and compare them with the time integration of model equations for cases discussed in table~\ref{tab: simulation_cases}. All the DNS results are non-dimensionalized according to scaling relations discussed in section~\ref{sec: theory}. 

\subsection{Dissipation in flat wall pores}
\begin{figure}[!t]
 \centering
 \includegraphics[width=\textwidth]{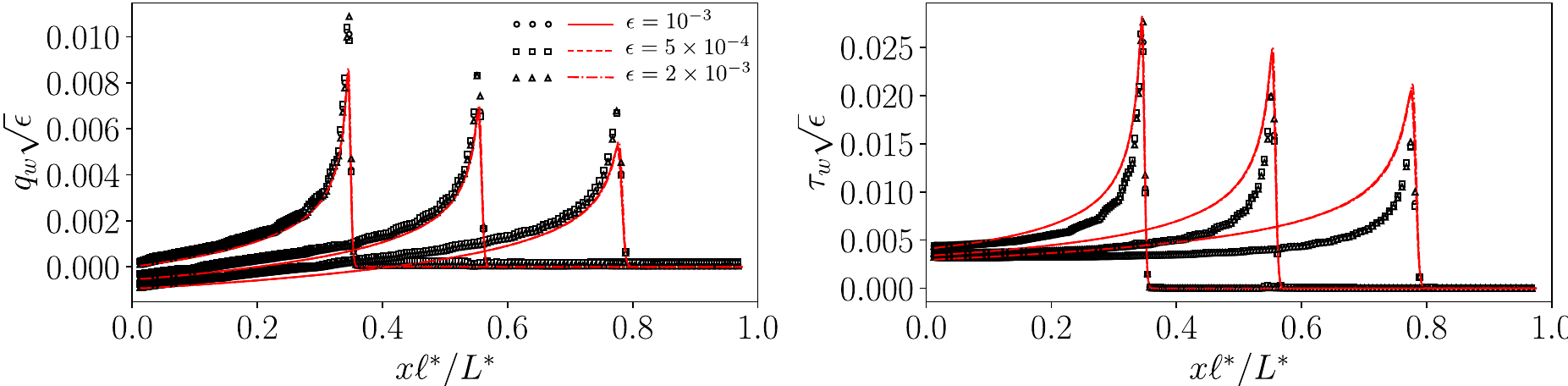}
 \put(-385,100){$(a)$}
 \put(-190,100){$(b)$}
 \caption{Variation of the normalized dimesionless $(a)$ wall-normal heat-flux $q_w$ and $(b)$ shear $\tau_w$ in $x$ at the no-slip isothermal wall $y=b(x)$ for flat wall pores $k=0$ (cases $1-3a$ in Table~\ref{tab: simulation_cases}) as the shock propagates inside the viscous pore at dimensionless time $t\ell^*/L^* = 0.35, 0.52, 0.69$. Nonlinear model : lines, DNS : markers}
 \label{fig: flat_pore_q_tau}
\end{figure}
\begin{figure}[!b]
 \centering
 \includegraphics[width=0.84\textwidth]{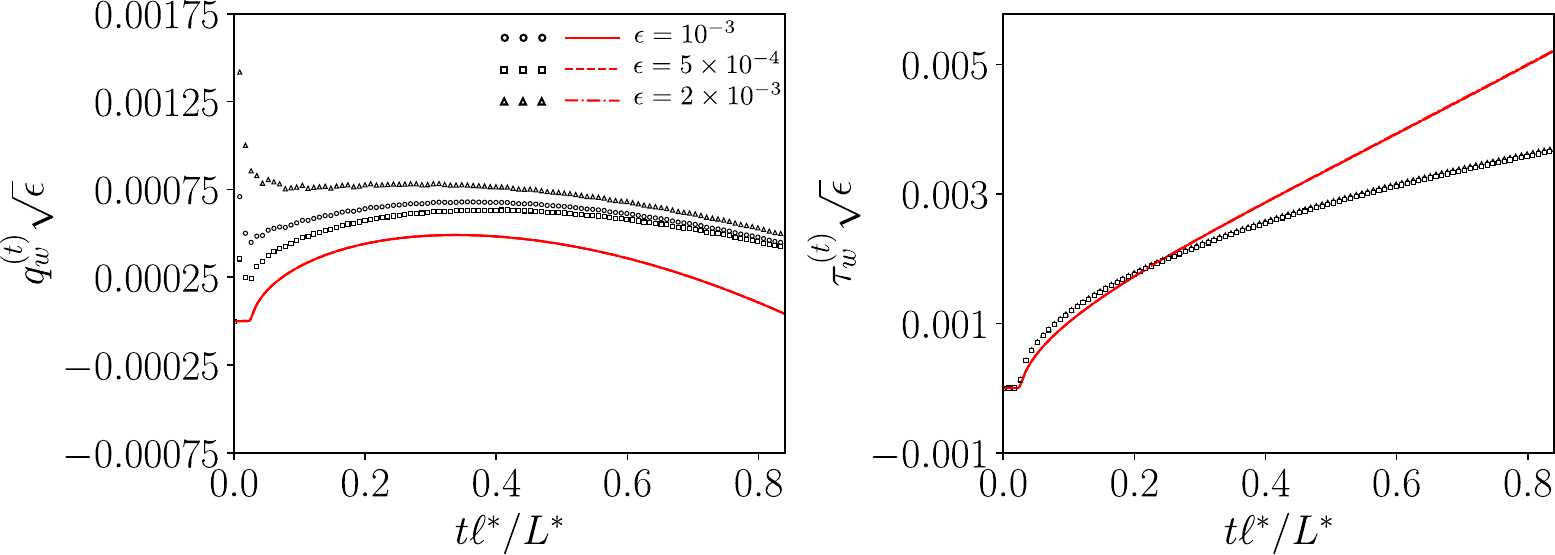}
  \put(-320,120){$(a)$}
 \put(-155,120){$(b)$}
 \caption{Variation of the normalized dimesionless wall-averaged total $(a)$ wall-normal heat-flux $q^{(t)}_w$ and $(b)$ shear $\tau^{(t)}_w$ in $t$ at the no-slip isothermal wall $y=b(x)$ for flat wall pores $k=0$ (cases $1-3a$ in Table~\ref{tab: simulation_cases}) as the shock propagates inside the viscous pore. Nonlinear model : lines, DNS : markers}
 \label{fig: flat_pore_q_tau_total}
\end{figure}
Figure~\ref{fig: flat_pore_y-avgd} shows the propagation of $y$-averaged pressure $\overline{p}'$ and $x$ direction velocity $\overline{u}'$ as the wave propagates inside the viscous duct, as calculated using the nonlinear model and the 2D DNS for case $1a$ in table~\ref{tab: simulation_cases}. In figures~\ref{fig: flat_pore_y-avgd}$c-h$, cross-sectional variation of the $x$ direction velocity and temperature perturbations are shown. As the shock propagates inside the viscous pore, $u'$ and $T'$ vary impulsively in $y$ and then relax due to both viscosity and thermal conduction, respectively.

Figure~\ref{fig: flat_pore_q_tau} shows the variation of the dimensionless wall-normal heat flux $q_w$ and shear $\tau_w$ for case $1a$ in table~\ref{tab: simulation_cases}, each defined as, 
\begin{equation}
 q_w = -\nabla T'\cdot\hat{\bfn},~~\tau_w = -\hat{\bft}\cdot\nabla\bfu'\cdot\hat{\bfn},~~
\end{equation}
where $T'$ is the dimensionless temperature perturbation, $\bfu'$ is the dimensionless velocity perturbation, and $\hat{\bfn}$ and $\hat{\bft}$ are outward unit normal and unit tangent vectors to the walls, respectively. As the shock passes, the temperature behind the shock rises which results in a strong outward heat flux in the walls, also known as shock-heating~\cite{rott1975thermally}. Away from the shock, the heat flux decays and eventually becomes negative indicating a local heating effect on the fluid. Due to the base temperature gradient which follows from $T_w(x)$, fluid at a lower temperature is streamed closer to the walls at a higher temperature, which results in the local heating of the fluid. Negative outward heat flux $q_w<0$ far away from the shock wave in figure~\ref{fig: flat_pore_q_tau}$a$ also indicates the heating of the fluid due to the walls. However, shear on the fluid always acts opposite to the flow and a thin boundary layer is formed behind the shock wave. Far from the shock wave, shear on the wall is positive $\tau_w > 0$ (see figure~\ref{fig: flat_pore_q_tau}$b$), denoting the skin-friction drag due to the boundary layer. Furthermore, figures~\ref{fig: flat_pore_q_tau}$a$ and $b$ also show that the dimenionless heat flux and the wall shear scale as $1/\sqrt{M_s - 1}$ for a weak shock wave of Mach number $M_s$, for constant shock-thickness.

Figure~\ref{fig: flat_pore_q_tau_total} shows the time evolution of the normalized dimensionless wall-averaged total heat-flux $q^{(t)}_w$ and shear $\tau^{(t)}_w$ at the wall $y=b(x)$ defined as, 
\begin{equation}
 q^{(t)}_w = \frac{1}{S}\int^S_0 q_w ds,~~\tau^{(t)}_w = \frac{1}{S}\int^S_0 \tau_w ds,
\end{equation}
where $s$ represents the path-length coordinate along the wall $b(x)$ and $S$ is the total length of the wall (surface area in 3D). Since the shear is always positive, total shear along the wall always increases in time as the wave propagates along the pore. However, since the wall heat-flux becomes negative (heat transfer inside the fluid) due to the temperature gradient, total wall heat-flux reaches a maximum and starts decreasing. When the combined effect of heat transfer inside the fluid dominates the viscous dissipation, thermoacoustic instability ensues (usually for very large $\ell^*$ or in the long wavelength limit~\cite{rott1969damped, gupta_lodato_scalo_2017}).

\begin{figure}[!b]
 \centering
 \includegraphics[width=\textwidth]{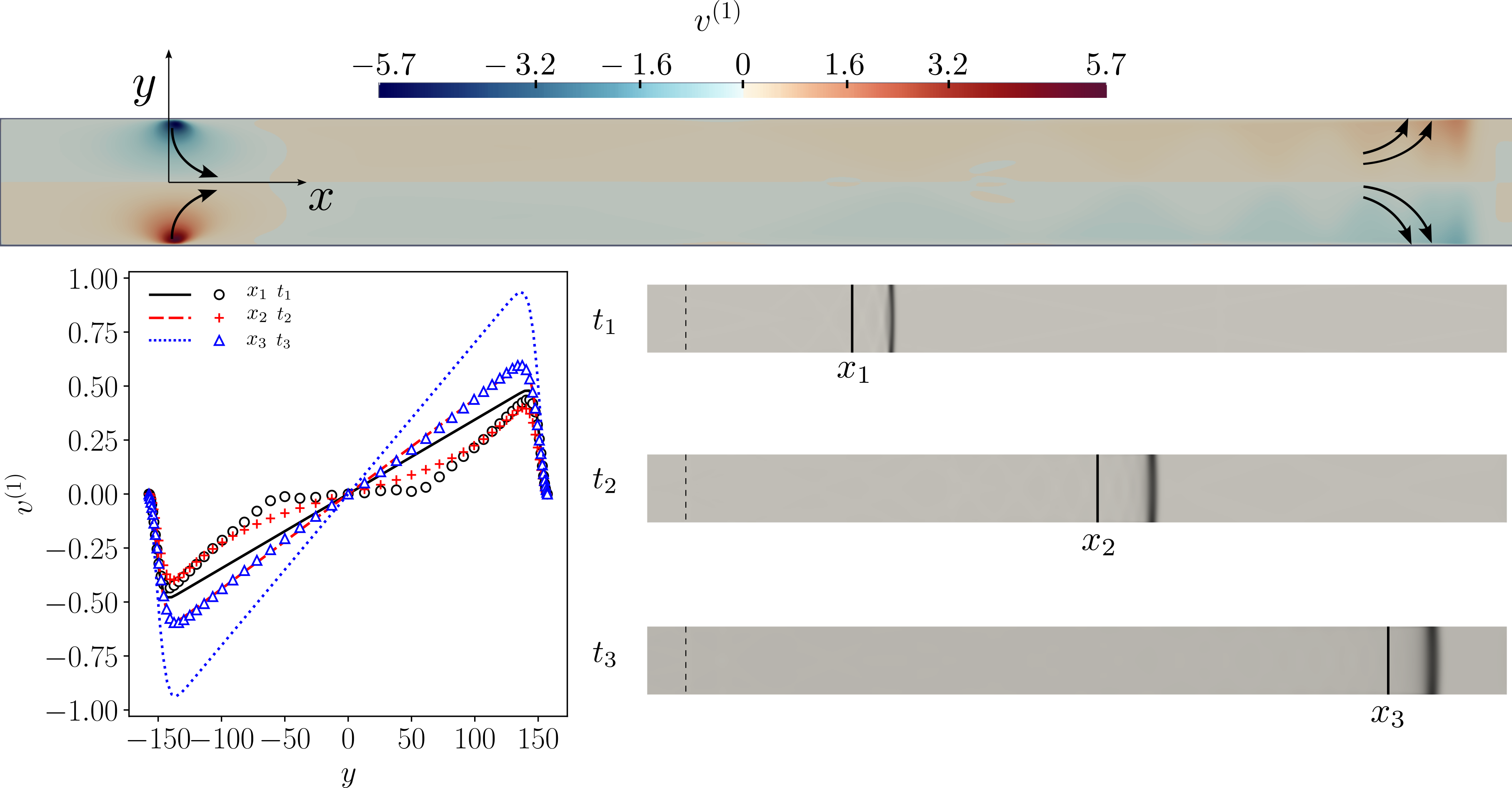}
  \put(-385,180){$(a)$}
 \put(-385,130){$(b)$}
  \put(-235,130){$(c)$}
 \caption{$(a)$ Contours of scaled $y$ velocity $v^{(1)}$ obtained from DNS for case $1a$ in Table~\ref{tab: simulation_cases} when the shock wave is at the middle of the viscous region, $(b)$ comparison of the captured $y$ velocity perturbations $v^{(1)}$ as obtained from the nonlinear model integration (lines) with the data obtained from the DNS (markers), plotted at $x_1=50, x_2=110, x_3=180$ at $t_1 = 51.8, t_2=105.3, t_3=157.2$, corresponding to $x$ locations behind the shock as schematically shown in $(c)$. We use the magnitude of divergence of velocity to visualize shock wave. }
 \label{fig: y-velocity}
\end{figure}
\subsection{Wall-normal velocity}
Figure~\ref{fig: y-velocity} shows the variation of $y$ velocity $v^{(1)}$ as obtained from the model equations and the DNS. As the wave enters the viscous region, flow is directed away from the no-slip isothermal walls resulting in $y$ direction acceleration of the flow. As the wave propagates in the viscous region, the local expansion of the gas behind the shock wave generates $v^{(1)}$ directed towards the walls. We note here that the \eqref{eq: v_1} for $v^{(1)}$ is derived using the continuity and the temperature equation. Thus, any $y$ direction acoustic waves (which can be captured only by combining the continuity and the unsteady $y$ momentum equation) are not captured in the nonlinear model \eqref{eq: u_bar_nl} and \eqref{eq: p_bar_nl}. Figure~\ref{fig: y-velocity} shows that the model captures the approximate variation of $v^{(1)}$ behind the shock wave. In the DNS, spatial oscillations of $v^{(1)}$ can be observed, which were not seen to be captured by the nonlinear model. 
Since the model captures the overall trend and order of magnitude of the $y$ direction velocity in the viscous region, the overall effects of the shear and the heat flux away from the walls are also captured by the model (c.f.~\eqref{eq: v_1}). 
Below, we discuss the effects of the wall waviness on the shear and the wall heat-transfer and the effective dissipation of the nonlinear waves due to the waviness. Furthermore, due to wavy walls, $y$ direction acoustic waves are also excited. Besides spatial variation of $v^{(1)}$, these acoustic waves do not interfere in the nonlinear acoustic wave dissipation mechanism. Hence, we outline only the wall shear and heat transfer effects of the wavy walls. 

\subsection{Effect of wall waviness}
\begin{figure}[!t]
 \centering
 \includegraphics[width=\textwidth]{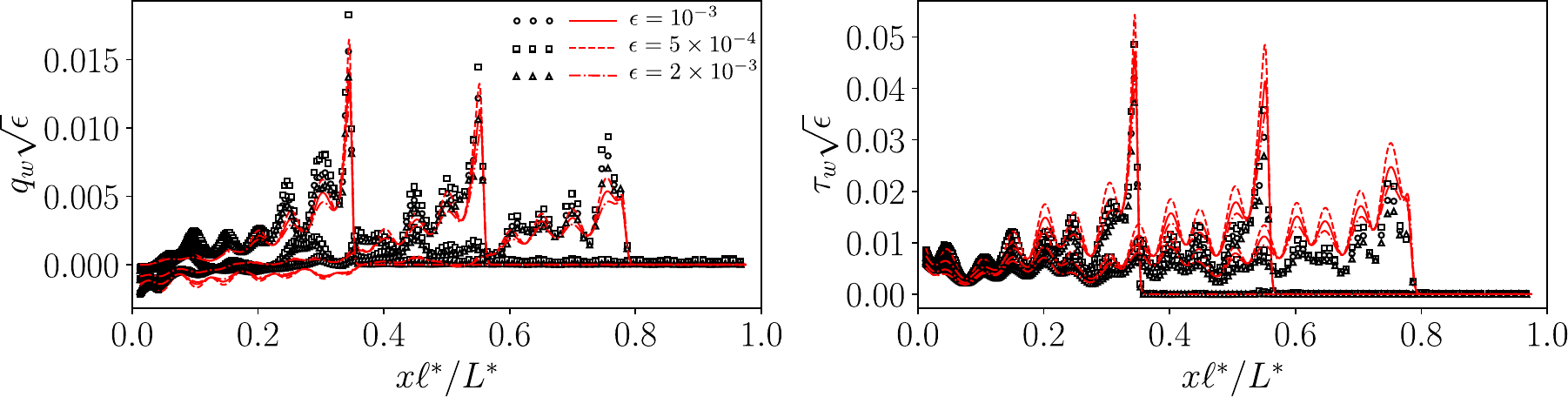}
  \put(-385,100){$(a)$}
 \put(-187,100){$(b)$}
 \caption{Variation of the normalized dimesionless $(a)$ wall-normal heat-flux $q_w$ and $(b)$ shear $\tau_w$ in $x$ in $x$ at the no-slip isothermal wall $y=b(x)$ as the shock propagates inside the viscous pore for cases $1-3b$ in Table~\ref{tab: simulation_cases} at dimensionless time $t\ell^*/L = 0.35, 0.52, 0.69$. Nonlinear model : lines, DNS : markers}
 \label{fig: qtau_k10}
\end{figure}
\begin{figure}[!b]
 \centering
 \includegraphics[width=0.84\textwidth]{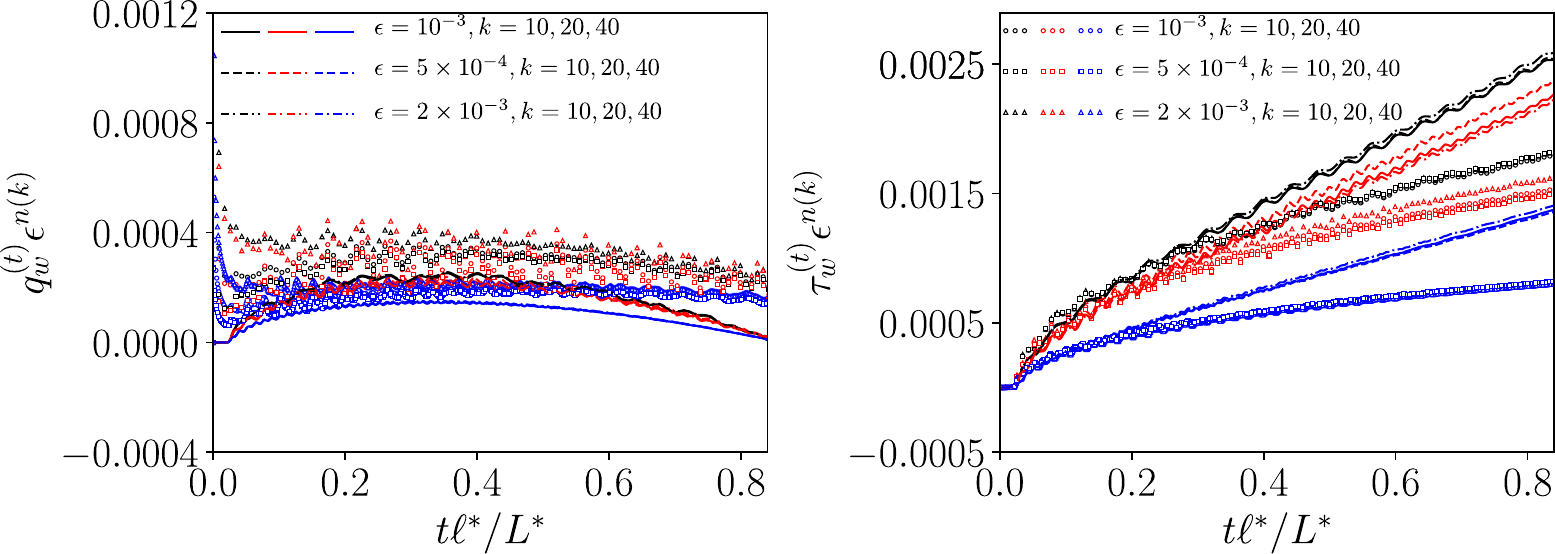}
   \put(-320,120){$(a)$}
 \put(-155,120){$(b)$}
 \caption{Variation of the normalized dimesionless wall-averaged $(a)$ total heat-flux $q_w$ and $(b)$ shear $\tau_w$ in $t$ at the no-slip isothermal wall $y=b(x)$ as the shock propagates inside the viscous pore for cases $1-3b, 1-3c,$ and $1-3d$ in Table~\ref{tab: simulation_cases}. Nonlinear model : lines, DNS : markers}
 \label{fig: qtau_k_comparison}
\end{figure}
Due to the waviness of the walls, the wall heat-transfer and shear fluctuate behind the shock wave after reaching a maximum as shown in figure~\ref{fig: qtau_k10}. The fluctuations are caused by oscillations of the wall normal $\hat{\bfn}$, which increase as the wavenumber $k=L^*/\lambda^*$ of the wall increases. We also note that the maximum values of the heat transfer and shear also oscillate due to the oscillatory contribution of the $x$ component of $\hat{\bfn}$. From figure~\ref{fig: qtau_k10}, we also note that the scaling of the dimensionless shear and wall heat-flux deviates from $\sim 1/\sqrt{\epsilon}$ for wavy walls.
Figure~\ref{fig: qtau_k_comparison} shows the variation of the average heat transfer and shear with the wall-waviness wavenumber $k$. As the walls become more wavy (increasing $k$), the total heat transfer and shear increase. Furthermore, the effect of waviness wavenumber is different for different strength parameters $\epsilon$.  The total heat flux and shear can be considered to scale as $\sim \epsilon^{-n(k)}$ where the scaling exponent $n=0.5$ for $k=0$ i.e. the flat wall pores. We note that as the wavenumber $k$ increases, $n(k)$ also increases with approximate values $n(10) \approx 0.65$, $n(20)\approx 0.75$, and $n(40)\approx 0.95$. Such a variation can be explained accounting for the variation in thickness of the wavefront due to thermoviscous dissipation. As the wave propagates inside the pore, the wavefront thickness ($\ell^*$) increases. Different waviness wavenumbers $k$ become comparable to the wavefront thickness at different locations within the pore, modifying the overall dependence of the wall shear and heat flux on the shock-strength parameter. We further note that dependence on waviness is not only due to $n(k)$. The wall shear and heat transfer scaled with $\epsilon^{n(k)}$ decrease for increasing $k$.

The discrepancy between the nonlinear model and the DNS varies for a wavy wall pore similar to a flat wall pore. We note that while the heat transfer is under-predicted in the nonlinear model, the shear is over-predicted. These seemingly nullifying effects result in a good match between the $y$-averaged pressure and velocity profiles. As shown in~\cite{gupta_lodato_scalo_2017}, the wall heat-transfer results in thermoacoustic amplification of the acoustic waves due to thermoacoustic overstability. For nonlinear acoustic waves, the total heat transfer out of the walls is underpredicted, indicating overprediction of thermoacoustic amplification in the nonlinear acoustic waves, which was also observed in~\cite{gupta_lodato_scalo_2017}.







\section{Conclusions}
\label{sec: conc}
In this work, we have derived and evaluated a nonlinear wave propagation model for the finite-amplitude nonlinear acoustic waves in narrow thermoviscous pores, typically encountered in thermoacoustic and aeroacoustic applications. We assumed the thickness of the wave $\ell^*$ inside the thermoviscous pore to scale inversely with the strength of the wave and derived the model in the stretched coordinates. We performed numerical simulations of the nonlinear wave propagation model equations and evaluated the results against the shock-resolved 2D DNS using a spectral difference code for the flat wall and the wavy wall pore geometries. For the flat-wall pores, we showed that the $y$-averaged pressure and velocity wave profiles obtained from the model equations match very well with the DNS. Furthermore, the dimensionless heat transfer and wall shear scale as $1/\sqrt{\epsilon}$ suggesting the dimesional heat transfer and wall shear to increase linearly with the shock strength $\epsilon$. We note that the scaling may change for wavefronts with the same shock strength but different $\ell^*$ ($\alpha$ in \eqref{eq: sim_init_p}).

For the pores with wavy walls, we observed that $y$-averaged pressure and velocity profiles exhibit oscillations at the length scale corresponding to the wall wavenumber $k$. Furthermore, the wall heat-transfer and shear also oscillate along the wall. The matching 
between the model equations and the DNS is reasonable for the wavy-wall cases. Additionally, the total (wall-averaged) heat transfer and shear evolve with a similar trend. We also showed that the total (wall-averaged) dimensionless heat-transfer and shear scale with the shock strength as $\epsilon^{-n(k)}$ where $n(k)\approx 0.65, 0.75, 0.95$ for $k=10, 20, 40$, respectively. Furthermore, the scaled total heat-transfer and shear increase with decreasing $k$.
The model equations for wavy-wall pores are useful in deriving computationally fast reduced order models for complex geometric shapes such as TPMS based heat exchangers and porous walls used in aeroacoustics. Current model equations and results will be utilized in formulating such models in future work. 







{We acknowledge the financial support received from Science and Engineering Research Board (SERB), Government of India under Grant No. SRG/2022/000728. Partial support from IITD under MFIRP Grant No. MI03600 is also acknowledged. We thank IIT Delhi HPC facility for computational resources. We also thank Mr. Altaf Ahmed for performing the validation study of the 2D Navier-Stokes solver. }


\appendix
\section{Solution of $\mathcal{O}(1)$ equations using Laplace transform}
Laplace transformed velocity and temperature perturbations are given as, 
\begin{align}
&\hat{u} = -\frac{1}{sf}\pfrac{\hat{p}}{x}\left(1 - \frac{\cosh\left(\eta y\right)}{\cosh\left(\eta b\right)}\right),\\
&s\hat{T} = \left(\frac{K}{s}\pfrac{\hat{p}}{x} + s(\gamma - 1) h \hat{p}\right)\left(1 - \frac{\cosh\left(\eta\sqrt{\mathrm{Pr}}y\right)}{\cosh\left(\eta\sqrt{\mathrm{Pr}}b\right)}\right) - \frac{\mathrm{Pr} K}{s}\pfrac{\hat{p}}{x}\left(1 - \frac{\cosh\left(\eta y\right)}{\cosh\left(\eta b \right)}\right),
\end{align}
where 
\begin{equation}
 K = \frac{h}{1-\mathrm{Pr}}\frac{dh}{dx},~\eta = \frac{\sqrt{s}}{h}.
\end{equation}
Consider the following inverse Laplace transform,
\begin{equation}
 f(t,a) = \mathcal{L}^{-1}\left(\frac{1}{s}\frac{\cosh\left(a\sqrt{s}y\right)}{\cosh\left(a\sqrt{s}b\right)}\right) = \sum \mathrm{Res}\left(\frac{e^{st}}{s}\frac{\cosh\left(a\sqrt{s}y\right)}{\cosh\left(a\sqrt{s}b\right)}\right) = \sum \mathrm{Res} F(s).
\end{equation}
All the poles of the function $F(s)$ are simple and are given by, 
\begin{equation}
 s =0, ~ s_n = -\left(n+\frac{1}{2}\right)\frac{\pi^2}{a^2b^2} = -\frac{z^2_n}{\left(ab\right)^2}.
\end{equation}
Computing the residues, we obtain, 
\begin{equation}
  f(t,a) = 1 + \sum^{\infty}_{n=0}\left(\frac{e^{s_n t}}{s_n}\frac{\cosh\left(iz_n\frac{y}{b}\right)}{\frac{d}{ds}\left(\cosh\left(ab\sqrt{s}\right)\right)|_{s_n}}\right) = 1 + \sum^{\infty}_{n=0}\left(\frac{a^2b^2e^{\frac{-z^2_n t}{ab}}}{-z^2_n}\frac{\cos\left(z_n\frac{y}{b}\right)}{\sinh\left(iz_n\right)\frac{ab}{2\sqrt{s_n}}}\right),
\end{equation}
which yields,
\begin{equation}
  f(t,a) = 1 + 2\sum^{\infty}_{n=0}\left(-1\right)^{n+1}\frac{e^{\frac{-z^2_n t}{ab}}}{z_n}\cos\left(z_n \frac{y}{b}\right) = 1 + \mathcal{G}(t,a).
\end{equation}
Inverting $\hat{u}$, we obtain, 
\begin{equation}
 u' = \frac{1}{f}\int^t_0\mathcal{G}(t-\tau, 1/h)\pfrac{p'}{x}(\tau)d\tau,
\end{equation}
where 
\begin{equation}
 \mathcal{G}(t,a) = 2\sum^{\infty}_{n=0}\left(-1\right)^{n+1}\frac{e^{\frac{-z^2_n t}{ab}}}{z_n}\cos\left(z_n \frac{y}{b}\right)~~\mathrm{where}~~ z_n = \left(n + \frac{1}{2}\right)\pi.\label{eq: relaxation_kernel_app}
\end{equation}
We also note that it is simpler to invert for $s\hat{T}$ as, 
\begin{align}
 \frac{\partial T'}{\partial t} = K\int^t_0\mathcal{G}(t-\tau, \sqrt{\mathrm{Pr}}/h)\pfrac{p'}{x}(\tau)d\tau &+ (\gamma - 1)h\int^t_0\mathcal{G}(t-\tau, \sqrt{\mathrm{Pr}}/h)\pfracTwo{p'}{t}(\tau)d\tau \nonumber\\
 &-\mathrm{Pr}K \int^t_0\mathcal{G}(t-\tau, 1/h)\pfrac{p'}{x}(\tau)d\tau.
\end{align}

\section{derivation of nonlinear equations}
\label{sec: Appb}
Second order $x-$momentum equation is, 
\begin{align}
 f\pfrac{u'}{t} + \epsilon\left(fu'\pfrac{u'}{x} + v'\pfrac{u'}{y}\right) = -\left(1 - \frac{\epsilon \rho'}{f}\right)\pfrac{p'}{x} + \left(1 - \frac{\epsilon \rho'}{f}\right)h\pfracTwo{u'}{y} + \nonumber\\
 \epsilon\left[\frac{4}{3}\frac{\partial}{\partial x}\left(h\pfrac{u'}{x}\right) + \frac{\partial}{\partial y}\left(T'\pfrac{u'}{y}\right) + \frac{\partial }{\partial y}\left(\frac{h}{3}\pfrac{v'}{x}\right) - \frac{2}{3}\frac{dh}{dx}\pfrac{v'}{y}\right].
\end{align}
Averaging in $y$ direction, and denoting, 
\begin{equation}
 \yavg{\phi} = \overline{\phi}, 
\end{equation}
we obtain,
\begin{align}
 \yavg{f\pfrac{u}{t}} + \yavg{\epsilon f\left(u'\pfrac{u'}{x} + v'\pfrac{u'}{y}\right)} = -\yavg{\left(1 - \frac{\epsilon \rho'}{f}\right)\pfrac{p'}{x}} + \nonumber \\
 \yavg{\left(1 - \frac{\epsilon \rho'}{f}\right)h\pfracTwo{u'}{y}} + \frac{4\epsilon}{3}\yavg{\frac{\partial}{\partial x}\left(h\pfrac{u'}{x}\right)}.
\end{align}
First term on LHS simplifies to 
\begin{equation}
 \yavg{f\pfrac{u}{t}} = f\pfrac{\overline{u}'}{t}.
\end{equation}
Let us look at the RHS before the second term on LHS. The first term on RHS is, 
\begin{equation}
 \yavg{\left(1 - \frac{\epsilon \rho'}{f}\right)\pfrac{p'}{x}} = \pfrac{\overline{p}'}{x} + \frac{\overline{p}'}{b}\frac{dp}{dx} - \yavg{\frac{\epsilon \rho'}{f}\pfrac{p'}{x}}.
\end{equation}
If we write, 
\begin{equation}
 p' = p^{(1)} + \epsilon p^{(2)}, 
\end{equation}
then we see that pressure is constant in $y$ at the leading order. Hence, 
\begin{equation}
 \overline{p} = p^{(1)} + \mathcal{O}(\epsilon)
\end{equation}
Hence, the third term above simplifies to
\begin{equation}
 \yavg{\frac{\epsilon \rho'}{f}\pfrac{p'}{x}} = \frac{\epsilon\overline{\rho}}{f}\pfrac{\overline{p}'}{x} + \mathcal{O}(\epsilon^2).
\end{equation}
The first term on RHS then becomes, 
\begin{equation}
 \yavg{\left(1 - \frac{\epsilon \rho'}{f}\right)\pfrac{p'}{x}} = \pfrac{\overline{p}'}{x} + \frac{\overline{p}'}{b}\frac{db}{dx} - \frac{\epsilon\overline{\rho}}{f}\pfrac{\overline{p}'}{x} + \mathcal{O}(\epsilon^2).
\end{equation}
The shear stress term can be simplified as, 
\begin{equation}
 \yavg{\left(1 - \frac{\epsilon \rho'}{f}\right)h\pfracTwo{u'}{y}} = \frac{h\tau_w}{b} - \epsilon h^2\yavg{\rho'\pfracTwo{u'}{y}}.
\end{equation}
Using by-parts, the above equation simplifies to, 
\begin{equation}
 \yavg{\left(1 - \frac{\epsilon \rho'}{f}\right)h\pfracTwo{u'}{y}} = \frac{h\tau_w}{b}\left(1 - \epsilon\gamma \overline{p}'\right) - \epsilon\yavg{\pfrac{T'}{y}\pfrac{u'}{y}} + \mathcal{O}(\epsilon^2)
\end{equation}
The bulk viscous stress term simplifies to, 
\begin{align}
 \frac{4\epsilon}{3}\yavg{\frac{\partial}{\partial x}\left(h\pfrac{u'}{x}\right)} = \frac{4\epsilon}{3}\left(\frac{\partial}{\partial x}\yavg{h\pfrac{u'}{x} }+ \frac{1}{b}\frac{db}{dx}\yavg{h\pfrac{u'}{x} }
 \right)\nonumber\\
 =\frac{4\epsilon}{3}\left(\frac{\partial }{\partial x} + \frac{1}{b}\frac{db}{dx}\right)\left(h\pfrac{\overline{u}'}{x} + \frac{h\overline{u}'}{b}\frac{db}{dx}\right)
\end{align}
The $v'$ term can be simplified as, 
\begin{equation}
 \epsilon\yavg{v'\pfrac{u'}{y}} = -\epsilon\yavg{u'\pfrac{v'}{y}}.
\end{equation}
Substituting from the $p$ equation, 
\begin{equation}
 -\epsilon\yavg{u'\pfrac{v'}{y}} = \epsilon\yavg{u'\left(\pfrac{p'}{t} + \pfrac{u'}{x}- \frac{h}{\mathrm{Pr}}\pfracTwo{T'}{y}\right)} + \mathcal{O}(\epsilon^2).
\end{equation}
Since pressure is constant in $y$ at $\mathcal{O}(\epsilon)$, we have, 
\begin{equation}
 -\epsilon\yavg{u'\pfrac{v'}{y}} =  \epsilon\yavg{u'\pfrac{u'}{x}} - \frac{\epsilon h}{\mathrm{Pr}}\yavg{u'\pfracTwo{T'}{y}} + \epsilon \overline{u}\pfrac{\overline{p}'}{t}.
\end{equation}
Substituting again for $\pfrac{\overline{p}'}{t}$, 
\begin{equation}
 \epsilon\yavg{v'\pfrac{u'}{y}} = \epsilon\yavg{u'\pfrac{u'}{x}} - \frac{\epsilon h}{\mathrm{Pr}}\yavg{u'\pfracTwo{T'}{y}} - \epsilon\overline{u}\left(\pfrac{\overline{u}'}{x} + \frac{\overline{u}'}{b}\frac{db}{dx} - \frac{q_w}{b Pr}\right)
\end{equation}
The final averaged momentum equation in $x$ direction is, 
\begin{align}
 f\pfrac{\overline{u}'}{t} + \yavg{\epsilon f\left(u'\pfrac{u'}{x} + v'\pfrac{u'}{y}\right)} = -\pfrac{\overline{p}'}{x} - \frac{\overline{p}'}{b}\frac{db}{dx} + \frac{\epsilon\overline{\rho}}{f}\pfrac{\overline{p}'}{x} + \frac{h\tau_w}{b}\left(1 - \epsilon\gamma \overline{p}'\right) \nonumber \\
 - \epsilon\yavg{\pfrac{T'}{y}\pfrac{u'}{y}} + \frac{4\epsilon}{3}\left(\frac{\partial }{\partial x} + \frac{1}{b}\frac{db}{dx}\right)\left(h\pfrac{\overline{u}'}{x} + \frac{h\overline{u}'}{b}\frac{db}{dx}\right).
\end{align}
Using similar manipulations, the $y$-averaged pressure equation \eqref{eq: p_bar_nl} can be derived.
\bibliography{references}


\end{document}